\documentclass[aps,prl,amsmath,reprint,superscriptaddress,nofootinbib]{revtex4-1}
\usepackage{graphicx}
\usepackage{hyperref}

\DeclareMathOperator{\Tr}{Tr}
\renewcommand{\Im}{\operatorname{Im}}
\newcommand{\BigO}[1]{\ensuremath{\operatorname{O}\bigl(#1\bigr)}}

\begin{document}

\title{Fundamental limits to nanoparticle extinction}

\author{O. D. Miller}
\affiliation{Department of Mathematics, Massachusetts Institute of Technology, Cambridge, MA 02139}
\author{C. W. Hsu}
\affiliation{Department of Physics, Massachusetts Institute of Technology, Cambridge, MA 02139}
\affiliation{Department of Physics, Harvard University, Cambridge, MA 02138}
\author{M. T. H. Reid}
\affiliation{Department of Mathematics, Massachusetts Institute of Technology, Cambridge, MA 02139}
\author{W. Qiu}
\affiliation{Department of Physics, Massachusetts Institute of Technology, Cambridge, MA 02139}
\author{B. G. DeLacy}
\affiliation{U.S. Army Edgewood Chemical Biological Center, Research and Technology Directorate, Aberdeen Proving Ground, MD 21010}
\author{J. D. Joannopoulos}
\affiliation{Department of Physics, Massachusetts Institute of Technology, Cambridge, MA 02139}
\author{M. Solja\v{c}i\'{c}}
\affiliation{Department of Physics, Massachusetts Institute of Technology, Cambridge, MA 02139}
\author{S. G. Johnson}
\affiliation{Department of Mathematics, Massachusetts Institute of Technology, Cambridge, MA 02139}


\begin{abstract}
We show that there are shape-independent upper bounds to the extinction cross section per unit volume of randomly oriented nanoparticles, given only material permittivity.  Underlying the limits are restrictive sum rules that constrain the distribution of quasistatic eigenvalues.  Surprisingly, optimally-designed spheroids, with only a single quasistatic degree of freedom, reach the upper bounds for four permittivity values.  Away from these permittivities, we demonstrate computationally-optimized structures that surpass spheroids and approach the fundamental limits.
\end{abstract}

\maketitle

Many applications \cite{Loo2005,Peer2007,Boisselier2009,Anker2008, Chang2010,Ohmachi2012,Appleyard2007,Chang2011} employ disordered collections of particles to absorb or scatter light, and the extinction for a given total particle volume (for a dilute system in which multiple scattering is negligible) is determined by the total (scattering + absorption) cross-section  per unit volume $\sigma_{\textrm{ext}}/V$ of the individual particles \cite{Bohren1983,Qiu2012}. In this paper, we prove fundamental upper bounds on $\sigma_{\textrm{ext}}/V$ for small particles of any shape, we show that previous work on maximizing particle scattering \cite{Link1999,Jain2006,Zhu2011,Roman-Velazquez2011,Qiu2012} (including ``super-scattering'' \cite{Ruan2010,Ruan2011,Verslegers2012}) was a factor of six or more from these bounds, and we employ a combination of analytical results and large-scale optimization (``inverse'' design) to discover nearly optimal particle shapes. Most previous work in this area was confined to spheres \cite{Jain2006,Qiu2012,Ruan2011} or a few high-symmetry shapes \cite{Link1999,Ruan2010,Zhu2011,Roman-Velazquez2011,Verslegers2012}, whereas we optimize numerically over shapes with $\approx 1000$ free parameters (and prove our theorem for completely arbitrary shapes) over the visible spectrum, and we also consider coated multimaterial shapes. We find that the optimal $\sigma_{\textrm{ext}}/V$ is invariably obtained for subwavelength particles where absorption dominates and the quasistatic approximation applies. We can then apply a little-known eigenproblem formulation of quasistatic electromagnetism in terms of ``resonances'' in the permittivity $\epsilon$ (\emph{not} in the frequency $\omega$) \cite{Fuchs1975,Ouyang1989,GarciadeAbajo2002, Mayergoyz2005,Kellogg1929}, and we employ various sum rules of these resonances \cite{Fuchs1975,Apell1996,Fuchs1976} to derive a bound on the cross section. Surprisingly, very different optimized shapes (such as ellipsoids or ``pinched'' tetrahedra) exhibit nearly identical $\sigma_{\textrm{ext}}(\omega)$ spectra (greatly superior to non-optimized particles) once $\sigma_{\textrm{ext}}$ is averaged over incident angle, a result we can explain in terms of the quasistatic resonances. Finally, we explain how our bounds provide materials guidance in various wavelength regimes, with potential applications ranging from cancer therapy \cite{Loo2005,Peer2007,Boisselier2009} and plasmonic biosensors \cite{Lal2007,Anker2008,Chang2010,Ohmachi2012} to next-generation solar cells \cite{Atwater2010a} and optical couplers \cite{Lin2013}.

Some previous bounds on optical properties of dilute particle suspensions have been derived.   Purcell derived a sum rule limiting the integral over all frequencies of extinction by spheroids \cite{Purcell1969}.  The limit has been extended to a variety of materials and structures  \cite{Gustafsson2007,Sohl2007a,Sohl2007,Sohl2007b}, but it is geometry-dependent and difficult to apply as a general rule.  Alternatively, many authors have bounded the effective ``metamaterial'' permittivity of composite media \cite{Bergman1979,Bergman1981,Milton1981,Milton2002}, a related but not identical problem.  The methods presented here, applied to the effective permittivity of a lossless dielectric, are able to reproduce the well-known Hashin--Strikman bounds \cite{Hashin1963,Lipton1993} of composite theory.

\begin{figure}
\includegraphics[width=\linewidth]{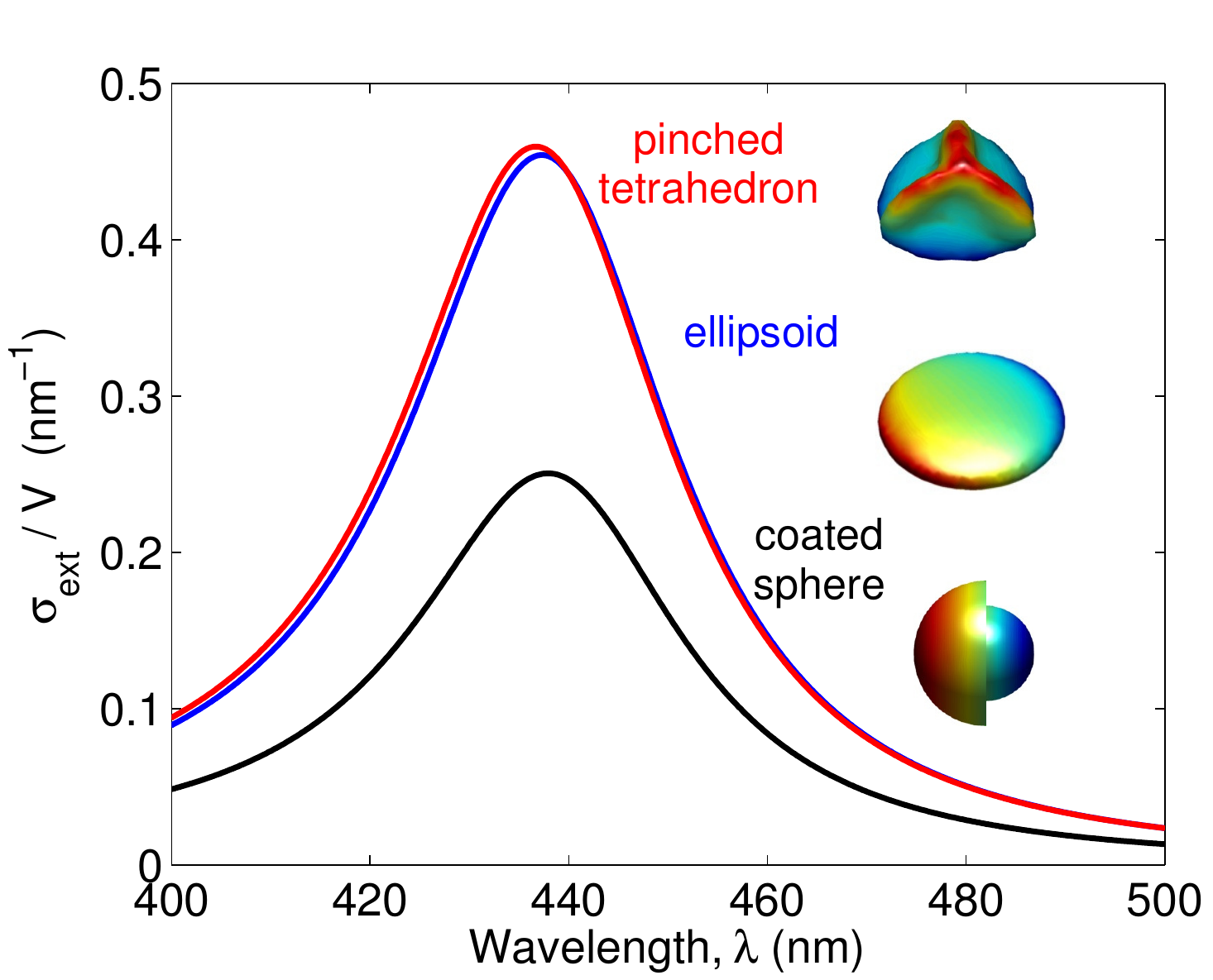}
\caption{\label{fig1}Angle-averaged extinction cross-section per unit volume of computationally-optimized Ag particles, designed for $\lambda_0=437$~nm and $\Delta\lambda=33$~nm. An ellipsoid provides almost twice the extinction of an optimally coated sphere, but optimizing over $\approx 1000$ spherical harmonics basis functions yields only a $2\%$ further improvement, due to fundamental limits on the eigenvalue distribution.  Surface coloring depicts the charge density on resonance, where $\epsilon_{Ag}(437\textrm{~nm}) \approx -5.5+0.7i$.  Particle dimensions are $\approx 10$~nm.}
\end{figure}

A single numerical optimization conceptually demonstrates many key findings for nanoparticle extinction.  To illustrate, we design a silver particle for maximum frequency-averaged extinction cross section per unit volume, $\sigma_{\textrm{ext}}/V$, over a $33nm$ bandwidth at center wavelength $\lambda=437nm$ ($Q=13$).  We do not impose quasistatic conditions \emph{a priori}; we employ the full Maxwell equations.  Ultimately, the optimizations always converged to very small, essentially quasistatic sizes.  

We employed a number of techniques to make the optimization tractable.  To quickly solve Maxwell's equations, we used a free-software implementation \cite{Reid,Reida} of the boundary-element method (BEM), which exploits piecewise constant media to express the scattering problem in terms of unknown fields on the surfaces only \cite{Harrington1993}.  Angle-averaging is essentially free with such a solver.  In many applications, the figure of merit is a frequency-averaged extinction, defined by the integral $\sigma_{\textrm{ext}} = \int \sigma_{\textrm{ext}}(\omega) H_{\Delta\omega}(\omega) d\omega$.   We efficiently compute this integral by contour integration, which for a Lorentzian $H$ of bandwidth $\Delta\omega$ reduces to a \emph{single} scattering problem at a \emph{complex} frequency $\omega_0+i\Delta\omega$ \cite{Hashemi2012,Liang2013}.  For optimization, the particle shape was parameterized by the zero level set of a sum of spherical harmonics \cite{Garboczi2002}, i.e. $r(\theta,\phi) = \sum_{lm}c_{lm} Y_{lm}(\theta,\phi)$ (restricting us to ``star-shaped'' structures).  Given the gradient of the objective with respect to these $\approx 1000$ degrees of freedom (efficiently computed by an adjoint method \cite{Strang2007,Jensen2011,Miller2012a}), we employ a free-software implementation \cite{Johnson} of standard nonlinear optimization algorithms \cite{Svanberg2002} to find a local optimum from a given starting point.  We also optimized the few degrees of freedom of coated spheres and ellipsoids for the sake of comparison.  

Figure~\ref{fig1} depicts the optimal particles and their respective extinction spectra.  The optimal designs were in the quasistatic limit, with dimensions $\approx 10nm$.  We see that uncoated ellipsoids provide significant gains over coated spheres, which already provide a substantial response \cite{Qiu2012,Loo2005,Boisselier2009,Delacy2013} (coated ellipsoids showed no further benefit).  This suggests a somewhat general principle that tuning resonances by geometrical deformation rather than by coatings enhances performance.  Oblate (``pancake'') ellipsoids are superior to prolate (``rod'') ellipsoids, because they couple to two of the three polarizations of randomly-oriented incident waves, as opposed to one.  In the much larger spherical harmonics design space, the optimal structure turned out to be a ``pinched'' tetrahedron (PT), which can be conceptualized as pinching a sphere towards the four centroids of the faces of an inscribed tetrahedron.  Surprisingly, the much larger design space yielded a structure that was only $2\%$ better than the best ellipsoid.  The two structures have very different responses for a given incidence angle and polarization; only when averaged over angle and polarization do the responses become nearly identical.  Also shown in Figure~\ref{fig1} are the imaginary parts of the charge densities for resonant incident waves, explained below.  Intuitively, the ellipsoid and pinched tetrahedron are better than the coated sphere because the opposing surface charges have larger spatial separations for a given volume.

The nearly identical spectra for the spheroid and PT can be explained by a fundamental restriction on quasistatic eigenmodes, which are prevented from fully coupling to external radiation.  In the quasistatic limit, the incident field is locally constant and the response of the system is determined by induced charge densities at the surfaces, whether by free charges in metals or bound charges in dielectrics.  One can construct the fields from the homogeneous Green's functions of the induced surface charges $\sigma(\mathbf{x})$.  For a surface $S$, the integral equation for the charge density is \cite{Fuchs1975,Ouyang1989,GarciadeAbajo2002, Mayergoyz2005,Kellogg1929}:
\begin{align}
\Lambda \sigma(\mathbf{x}) - \underbrace{\int_S \hat{\mathbf{n}}(\mathbf{x}) \cdot  \mathbf{G}^E(\mathbf{x}-\mathbf{x'}) \sigma(\mathbf{x}')  \,\mathrm{d}S'}_{\hat{K}\sigma} = \mathbf{E}^{inc}(\mathbf{x}) \cdot \hat{\mathbf{n}}(\mathbf{x})
\label{eq:IntgrlEqn}
\end{align} 
where $\Lambda = (\epsilon_{\textrm{int}}+\epsilon_{\textrm{ext}})/2(\epsilon_{\textrm{int}}-\epsilon_{\textrm{ext}})$ relates interior and exterior permittivities, the electrostatic Green's function $\mathbf{G}^E(\mathbf{x}) = \mathbf{x}/4\pi|\mathbf{x}|^3$, and $\mathbf{E}^{inc}(\mathbf{x}) \cdot \hat{\mathbf{n}}(\mathbf{x})$ is the normal component of the incident field at $\mathbf{x}$.  As distinguished from the resonant \emph{frequencies} of Maxwell's equations, there are resonant \emph{permittivities} $\epsilon_{\textrm{int}}/\epsilon_{\textrm{ext}}$ for the quasistatic integral equation.  These are negative, real-valued permittivities $\epsilon_n$ at which self-sustaining charge densities exist \emph{without} external fields, for specific eigenmodes $\sigma_n$ satisfying $\hat{K} \sigma_n = \lambda_n \sigma_n$, where $\hat{K}$ is the Neumann-Poincar\'{e} integral operator defined by Eq.~(\ref{eq:IntgrlEqn}).  The eigenvalues $\lambda_n$ lie in the interval $\left[-1/2,1/2\right]$ \cite{Kellogg1929,Mayergoyz2005,Sandu2012}, such that $\epsilon_n < 0$.  The left-eigenvectors of $K$, denoted $\tau_n$, have the same eigenvalue spectrum as the $\sigma_n$ and provide the orthogonality condition $\left\langle \sigma_n, \tau_m \right\rangle = \int_S \sigma_n \tau_m \,\mathrm{d}S = \delta_{mn}$ \cite{Mayergoyz2005}.

The eigenmodes contribute to absorption and scattering through $\alpha$, the particle's polarizability per unit volume $V$, relating the incident field to the dipole moment by $p_{\ell} = V\sum_{m}\alpha_{\ell m}E_{m}^{inc}$.  The dipole moment is given from the surface charge density by $\mathbf{p} = \int_S \mathbf{x} \sigma dA$.  Decomposing the charge density as a superposition of eigenmodes, $\sigma = \sum_n c_n \sigma_n$, yields
\begin{align}
\alpha_{\ell m} = \sum_n \frac{ p_n^{\ell m} }{L_n - \xi(\omega)} 
\end{align}
where $p_n^{\ell m} = \left\langle \sigma_n, x_{\ell} \right\rangle \left\langle \tau_n, \hat{n}_{m} \right\rangle / V$ is the dipole strength of each mode, $L_n = 1/2 - \lambda_n$ is the depolarization factor, and $\xi(\omega) = -\epsilon_{\textrm{int}}/(\epsilon_{\textrm{int}} - \epsilon_{\textrm{ext}})$ represents the relative properties of the interior and exterior materials.

The distribution of eigenmodes, and therefore the induced susceptibility, is restricted by two crucial sum rules.  The first is the \emph{f-sum} rule \cite{Apell1996,SuppMat} limiting the total dipole strength:
\begin{align}
\sum_n p_n^{\ell m} = \delta_{\ell m}.
\label{eq:SumRule1}
\end{align}
The second sum rule \cite{Fuchs1976,SuppMat} states that the weighted average of the depolarization factors must be $1/3$:
\begin{align}
\left\langle L_n \right\rangle = \frac{\sum_n p_n L_n}{\sum_n p_n} = \frac{1}{3}
\label{eq:SumRule2}
\end{align}
where $p_n$ denotes $\sum_{\ell} p_n^{\ell\ell}$.
  
A sphere has a depolarization factor of $1/3$, leading to a ``plasmon'' resonance at $\epsilon \approx -2$ ($\xi=1/3$).  Eq.~(\ref{eq:SumRule2}) dictates that the average depolarization factor of \emph{every} structure must equal that of the sphere; roughly, the average resonant permittivity of every structure must also be $-2$.  Although it was exploited for composites with certain symmetries \cite{Bergman1978,Stroud1979}, this general property has not been widely recognized and is very important in limiting possible extinction rates. 
\begin{figure}
\includegraphics[width=\linewidth]{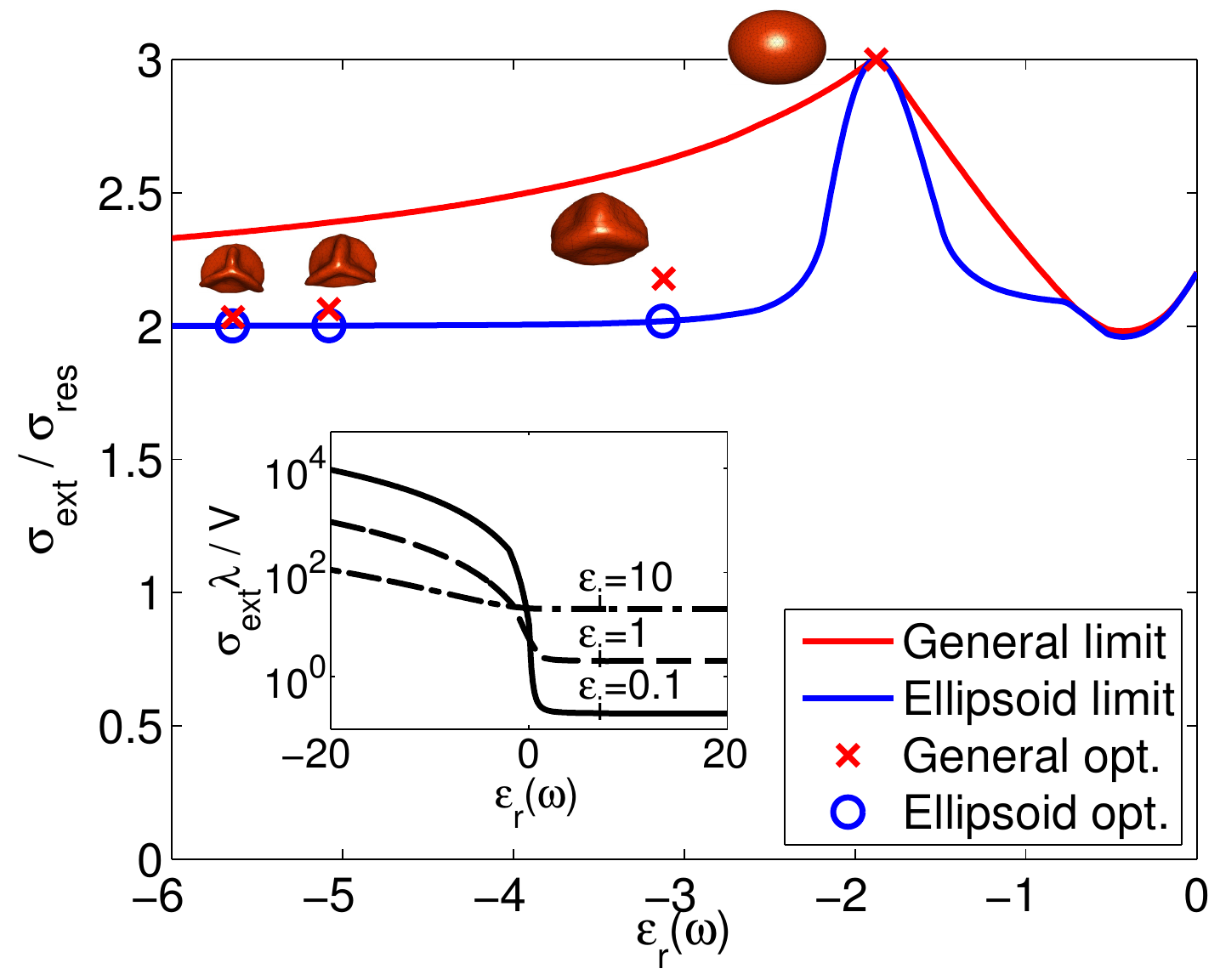}
\caption{\label{fig2}Fundamental extinction limits, normalized to the maximum extinction of a single-polarization resonance, $\sigma_{\textrm{res}}$.  Spheres are not optimal for absolute $\sigma_{\textrm{ext}}$ (see inset), but do enable full coupling to three polarizations, given by the normalized value $\sigma_{\textrm{ext}}/\sigma_{\textrm{res}}$.  Markers indicate computationally-optimized structures. $\epsilon_{\textrm{i}}(\omega)$ is taken to be that of Ag, although this has only a small effect on the lineshape. Ellipsoids can approach the general bounds in four limits: $\epsilon_{\textrm{r}}\rightarrow -\infty$ (oblate disk), $\epsilon_{\textrm{r}} = -2$ (sphere), $\epsilon_{\textrm{r}} = -1$ (cylinder), and $\epsilon_{\textrm{r}} = 0$ (oblate disk). Computationally-optimized “pinched tetrahedra”  improve upon ellipsoids at intermediate $\epsilon_{\textrm{r}}(\omega)$. Inset: upper bound on $\sigma_{\textrm{ext}}\lambda/V$, which increases with $\epsilon_r^2/\epsilon_i$ ($\epsilon_r<0$).}
\end{figure}

The average extinction of randomly-oriented particles is proportional to the imaginary part of $\Tr\alpha_{\ell m}$ \cite{Bohren1983}:
\begin{align}
\frac{\sigma_{\textrm{ext}}}{V} = \frac{2\pi}{3\lambda} \sum_n \Im \left[ \frac{1}{L_n-\xi(\omega)} \right] p_n
\label{eq:AvgExtinction}
\end{align}
A resonance occurs for $L_n=\xi_{\textrm{r}}(\omega)$, where $r$ and $i$ subscripts denote real and imaginary parts, respectively.  For particles in vacuum with susceptibility $\chi(\omega) = \epsilon(\omega) - 1$, $\xi(\omega)=-1/\chi(\omega)$.  Only metals, with $\epsilon_{\textrm{r}}(\omega)<0$, can achieve $0<\xi_{\textrm{r}}<1$, and therefore exhibit quasistatic surface-plasmon modes.  To maximize extinction, ideally a particle would have all of its dipole strength located exactly at $L_n=\xi_{\textrm{r}}(\omega)$.  However, the second sum rule guarantees that except in the case $\xi_{\textrm{r}} = 1/3$, this is not possible; there must always be a counter-balancing dipole moment such that $\left\langle L_n \right\rangle = 1/3$. 

For a given material parameter $\xi_{\textrm{r}}(\omega)$, we can show that the optimal distribution of eigenmodes has at most two distinct depolarization factors, $L_1$ and $L_2$.  We have rigorously derived the exact locations of the two eigenvalues \cite{SuppMat}, but for relevant materials a simple solution suffices:
\begin{align}
(L_1,L_2) = 
\begin{cases}
(\xi_{\textrm{r}},1) & 0 < \xi_{\textrm{r}} < 1/3 \\
(0,\xi_{\textrm{r}}) & 1/3 < \xi_{\textrm{r}} < 1 \\
(0,1) & \xi_{\textrm{r}}<0 \text{ or } \xi_{\textrm{r}}>1
\end{cases}
\label{eq:l1l2}
\end{align}
which corresponds to placing as much of the dipole moment as possible on resonance ($L=\xi_{\textrm{r}}$), and the rest of the dipole strength at the opposite boundary to satisfy the second sum rule.  Eq.~(\ref{eq:l1l2}) is exact for $\xi_{\textrm{i}}=0$ (both a low-loss $\chi_{\textrm{i}}=0$ and infinite-loss $\chi_{\textrm{i}}\rightarrow \infty$ limit), but is also very accurate (error $< 10^{-3}$) otherwise.  Distributing the dipole strength according to Eqs.~(\ref{eq:SumRule1},\ref{eq:SumRule2},\ref{eq:l1l2}) yields the upper limit to the extinction per unit volume:
\begin{align}
\frac{\sigma_{\textrm{ext}}}{V} \leq \frac{2\pi}{3\lambda} 
\begin{cases}
    \frac{ 2\chi_{\textrm{r}}^3(1+\chi_{\textrm{r}}) + \chi_{\textrm{i}}^2(3+2\chi_{\textrm{r}}+4\chi_{\textrm{r}}^2) + 2\chi_{\textrm{i}}^4 }{ \chi_{\textrm{i}} \left(\chi_{\textrm{i}}^2 + (1+\chi_{\textrm{r}})^2 \right) } & 0 < -\frac{\chi_{\textrm{r}}}{|\chi|^2} < \frac{1}{3} \\
    3\chi_{\textrm{i}} - \frac{\chi_{\textrm{r}}}{\chi_{\textrm{i}}}|\chi|^2 & \frac{1}{3} < -\frac{\chi_{\textrm{r}}}{|\chi|^2} < 1 \\
    \chi_{\textrm{i}} \left( 2 + \frac{1}{\chi_{\textrm{i}}^2 + (1+\chi_{\textrm{r}})^2} \right) & \textrm{else}
\end{cases}
\label{eq:BoundsChi}
\end{align}
which provides a limit for any possible susceptibility, independent of geometry.  Ideal scatterers are metals with very negative real permittivities and small imaginary permittivities; for $\epsilon_\textrm{i} \ll |\epsilon_\textrm{r}|$, Eq.~(\ref{eq:BoundsChi}) simplifies to:
\begin{align}
\frac{\sigma_{\textrm{ext}}}{V} \leq \frac{4\pi}{3\lambda}\frac{\epsilon_{\textrm{r}}^2}{\epsilon_{\textrm{i}}} + \BigO{\epsilon_i}
\label{eq:BoundsEps}
\end{align}
where the ``Big O'' notation indicates the asymptotic scaling of the higher-order term.
\begin{figure}
\includegraphics[width=\linewidth]{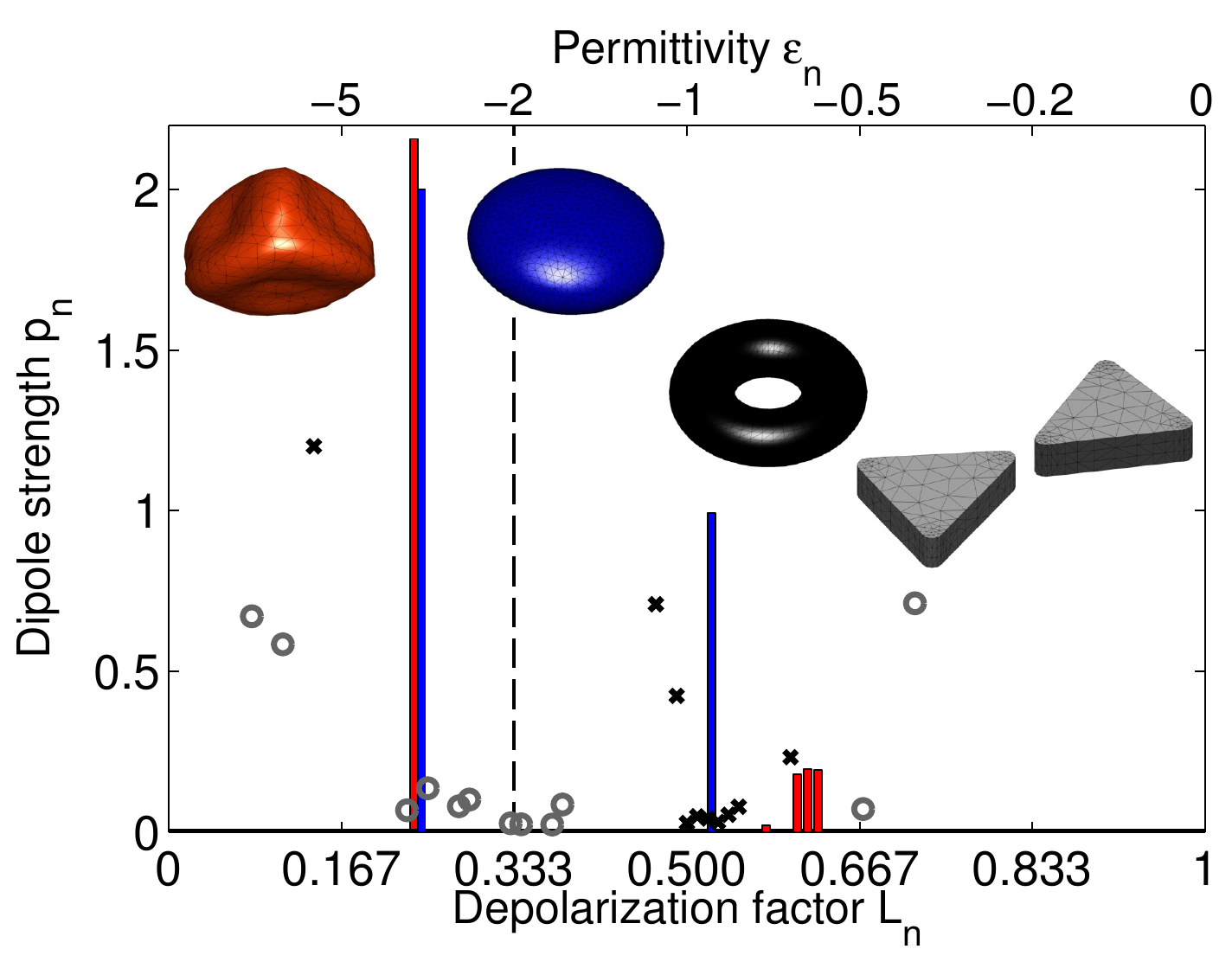}
\caption{\label{fig3}Extinction maximization is essentially an eigenvalue optimization problem.  The optimal pinched tetrahedron (PT) and ellipsoid have degenerate modes increasing the dipole strength at the optimization permittivity, in this case $\epsilon_{\textrm{r}}(\omega)=-3.2$.  The PT outperforms the ellipsoid because its ``undesirable'' eigenmodes are closer to $L=1$, enabling a larger dipole strength at $L_1=0.24$ ($\epsilon_1=-3.2$).  Equally important is the lack of other bright modes; Torii and bowtie antennas, for example, have disperse modes, reducing overall extinction.  The PT/ellipsoid modes coincide at $L_1=0.24$ but are split for visualization.}
\end{figure}

Eqs.~(\ref{eq:BoundsChi},\ref{eq:BoundsEps}) represent fundamental limits to quasistatic particle extinction.  Fig.~\ref{fig2} illustrates these limits by normalizing them relative to the value of extinction on resonance, $\sigma_{\textrm{res}}=2\pi/3\lambda\xi_{\textrm{i}}(\omega)$, and comparing them to ellipsoid limits computed through non-linear optimization \cite{Johnson}.  The structural eigenmodes were computed with BEM software \cite{Hohenester2012}.  $\sigma_{\textrm{ext}}/\sigma_{\textrm{res}}$ can be thought of as the number of fully coupled polarizations; only at $\xi_{\textrm{r}}=1/3$ ($\epsilon_{\textrm{r}}\approx-2$) can full coupling to all three polarizations occur.  Thus we see why ellipsoids perform very well, and why the optimal structure of Fig.~\ref{fig1} barely outperformed the ideal ellipsoid: in many cases, full coupling to two polarizations closely approaches the ideal performance.  This is exactly true for $\epsilon_{\textrm{r}} \rightarrow -\infty$, one of the cases in which ellipsoids reach the upper bound.  The other three cases are: $\epsilon_{\textrm{r}} = -2$, $\epsilon_{\textrm{r}} = -1$, and $\epsilon_{\textrm{r}} = 0$, for which a sphere, infinite cylinder, and infinitely thin disk are optimal, respectively.  In each case, the spheroid depolarization factors \cite{Bohren1983} are identical to those of the optimal general shape, given by Eq.~(\ref{eq:l1l2}). 

Included in Fig.~\ref{fig2} are optimizations at other permittivities (assuming the complex permittivity of Ag); we see that there is a family of ``pinched tetrahedron'' structures that emerge as superior design choices over ellipsoids, and approach the upper limits.  It is important to note that spheres are not globally optimal, as the normalization factor $\sigma_{\textrm{res}}$ is a function of $\epsilon_r$.  The inset of Fig.~\ref{fig2} shows the absolute extinction, which scales as $\epsilon_{\textrm{r}}^2/\epsilon_{\textrm{i}}$. 

Fig.~\ref{fig3} shows the depolarization factor distributions of the ideal pinched tetrahedron and ellipsoid structures, as well as non-ideal structures.  We see that the dipole moments are largely concentrated at the desired permittivity, except as required to keep the centroid of $L_n$ equal to $1/3$.  The tetrahedra have the off-resonance dipole moments distributed closer to the boundary $L_n=1$ than ellipsoids, explaining the slightly superior performance; to reach the upper bounds of Fig.~\ref{fig2} the off-resonance dipole strength would have to occur exactly at $L_n=1$.  

\begin{figure}
\includegraphics[width=\linewidth]{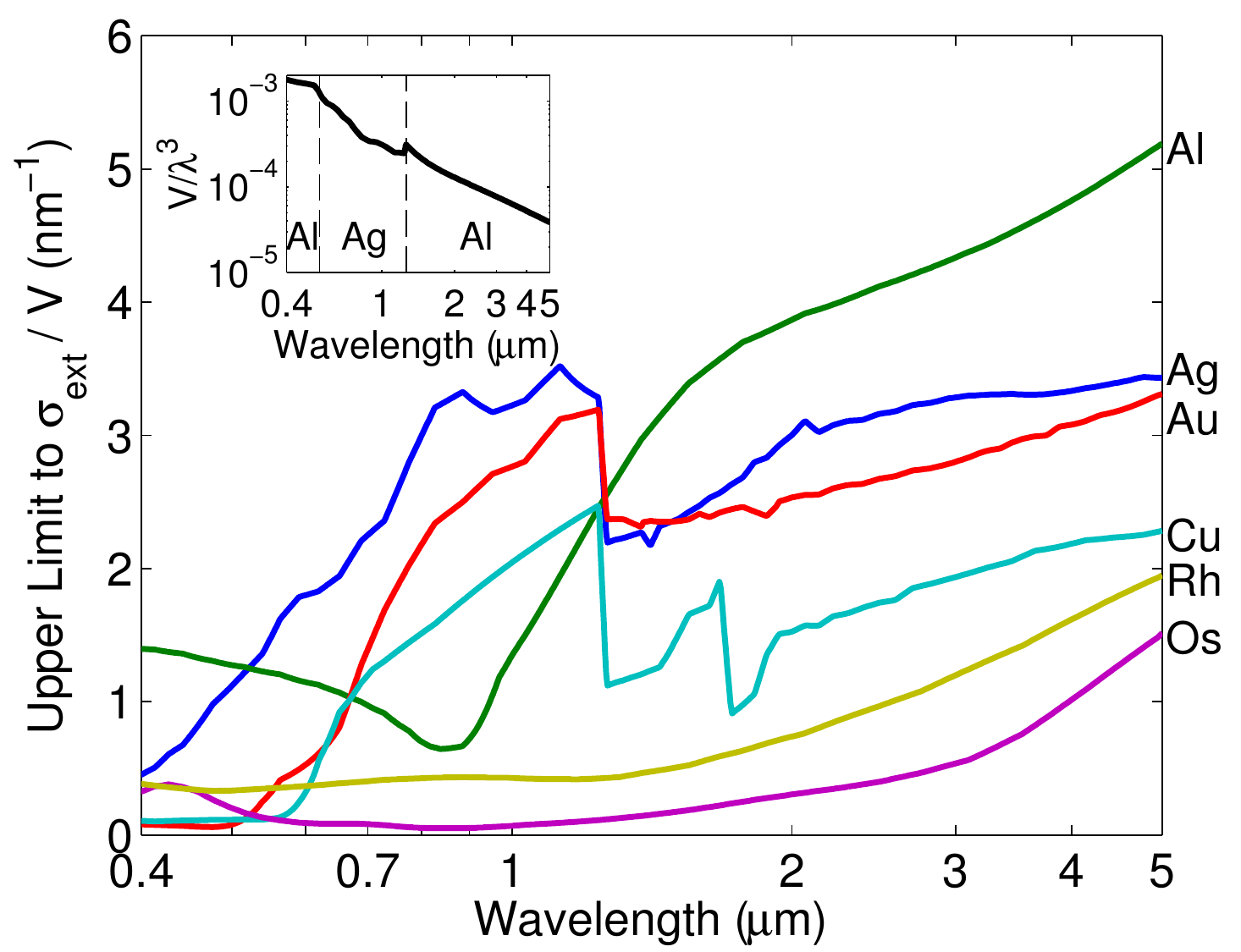}
\caption{\label{fig4}Shape-independent fundamental limit to extinction per unit volume for the highest-performing metals \cite{Palik1998a} at visible and infrared wavelengths. A mixture of Al and Ag nanoparticles, properly designed, could provide ideal extinction over the visible and near- to mid-infrared. Inset: minimum volume fraction, $V/\lambda^3$, required for $\sigma_{\textrm{ext}}=\lambda^2$. It is possible to achieve $\lambda^2$ cross-sections for $V/\lambda^3 < 10^{-3}$.}
\end{figure}

Fig.~\ref{fig4} illustrates the general utility of the bounds of Eq.~(\ref{eq:BoundsChi}).  For a given permittivity, a maximum extinction per unit volume can be computed independent of structure.  This has important implications for material selection, which varies by application and frequency.  Although the bounds are quasistatic, it may be that the quasistatic bound is optimal at any size (extinction likely cannot increase equally with $V$).  Indeed, the infrared extinction limits are three orders of magnitude larger than the best non-quasistatic particles investigated to date \cite{Appleyard2007}.  Although the bounds are for a single frequency, when used in conjunction with the material quality factors (known to be geometry-independent \cite{Wang2006a}), rational design for any bandwidth can be undertaken.

It is interesting to compare the structures presented here to ``super-scattering'' structures recently proposed in the literature \cite{Ruan2010,Ruan2011,Verslegers2012}.  Of primary importance is the figure of merit (FOM).  The extinction can be normalized by volume (as here), geometric cross section or squared wavelength.  For applications, however, it appears that volume or weight is the most relevant normalization.  Normalizing by $\lambda^2$, as in \cite{Ruan2010,Ruan2011,Verslegers2012}, favors larger particles approaching wavelength-scale.  A smaller particle with larger $\sigma_{\textrm{ext}}/V$ likely cannot extinguish a full square wavelength.  Yet a dilute mixture of such particles could, with much smaller volumes.  As an example, \emph{two} quasistatic nano-ellipsoids, with an 8:1 major to minor axis ratio can achieve the same $\sigma_{\textrm{ext}} /\lambda^2$ as the single particle in \cite{Ruan2011}, while requiring $1/270^{th}$ of the volume.  A single ``channel'' in a non-spherical structure can extinguish much more strongly than multiple channels in a spherical structure.

It is important to delineate the materials and structures for which Eqs.~(\ref{eq:BoundsChi},\ref{eq:BoundsEps}) are valid.  The materials are assumed to be linear, isotropic, and non-magnetic.    The eigenmode decomposition of $\hat{K}$, critical for our approach, requires that every surface have the same interior and exterior permittivities (i.e. scalar $\Lambda$).  Thus, our limits do not apply to layered structures or more than two permittivities, but are valid for arbitrarily many interacting objects of the same permittivity, with or without holes (e.g. torii).  Preliminary computational optimizations of coated structures have not found designs that can outperform the uncoated limits provided here; methods similar to those in Refs.~\cite{Roman-Velazquez2011,Ammari2013} may bound such structures.

Small, absorbing nanoparticles show promise for a variety of scientific and technical applications.  Experimentally approaching the limits derived here would already represent a significant achievement.  A possible further improvement could come from harnessing exotic material systems \cite{Wurthner2011,Delacy2013}, where geometry-dependent material resonances cannot be modeled with bulk permittivities.

This work was supported by the Army Research Office through the Institute for Soldier Nanotechnologies under Contract No. W911NF-07-D0004, and by the AFOSR Multidisciplinary Research Program of the University Research Initiative (MURI) for Complex and Robust On-chip Nanophotonics under Grant No. FA9550-09-1-0704.


\end{document}


\title{Supplementary Material for ``Fundamental limits to nanoparticle extinction''}

\author{O. D. Miller}
\affiliation{Department of Mathematics, Massachusetts Institute of Technology, Cambridge, MA 02139}
\author{C. W. Hsu}
\affiliation{Department of Physics, Massachusetts Institute of Technology, Cambridge, MA 02139}
\affiliation{Department of Physics, Harvard University, Cambridge, MA 02138}
\author{M. T. H. Reid}
\affiliation{Department of Mathematics, Massachusetts Institute of Technology, Cambridge, MA 02139}
\author{W. Qiu}
\affiliation{Department of Physics, Massachusetts Institute of Technology, Cambridge, MA 02139}
\author{B. G. DeLacy}
\affiliation{U.S. Army Edgewood Chemical Biological Center, Research and Technology Directorate, Aberdeen Proving Ground, MD 21010}
\author{J. D. Joannopoulos}
\affiliation{Department of Physics, Massachusetts Institute of Technology, Cambridge, MA 02139}
\author{M. Solja\v{c}i\'{c}}
\affiliation{Department of Physics, Massachusetts Institute of Technology, Cambridge, MA 02139}
\author{S. G. Johnson}
\affiliation{Department of Mathematics, Massachusetts Institute of Technology, Cambridge, MA 02139}

\maketitle

\section{Proof of optimal eigenvalue distribution}
We will discuss the optimal distribution of eigenvalues.   We emphasize that we are only discussing ``bright'' modes, i.e. eigenvalues with a non-zero dipole moment.  They are the only ones that contribute to extinction, and are the only ones subject to the sum rules.  We note again that the depolarization factor $L_n$ is related to the eigenvalue $\lambda_n$ by $L_n = 1/2 - \lambda_n$, so that in many instances the words can be used interchangeably.

First, we will show that the optimal distribution takes fewer than three distinct eigenvalues.  Intuitively, this can be thought of arising from the fact that the sum in Eq.~(5) is \emph{incoherent}.  There is no mixing of modal contributions, and aside from meeting the sum rule requirements, having more than one eigenvalue on the ``same side'' of $1/3$ cannot help.  One of the eigenvalues will be strictly better, or worse, than the other.

Let us assume that there are at least three distinct eigenvalues, with increasing depolarization factors $L_1<L_2<L_3$, and non-zero dipole moments $p_1$, $p_2$, and $p_3$.  We will show that the ideal distribution always yields at least one of the dipole moments to be zero.  The three eigenvalues need only be a subset of all of the eigenvalues (we will hold the other $p_i$ and $L_i$ fixed), such that the three satisfy modified sum rules
\begin{subequations}
\begin{align}
p_1+p_2+p_3 = c_1 \\
p_1L_1 + p_2L_2 + p_3L_3 = c_2
\end{align}
\end{subequations}
where $c_1 \leq 3$ and $c_2 \leq 1$.  

Assume first that both $L_1$,$L_2 \leq 1/3$.  Then we can set $p_1$ as an independent variable $x$ (i.e. $p_1=x$) in the interval $[0,\frac{c_1L_3-c_2}{L_3-L_1}]$; varying $x$ varies $p_2$ and $p_3$ accordingly:
\begin{subequations}
\begin{align}
p_2 = p_{2,\max} - x\frac{L_3-L_1}{L_3-L_2} \\
p_3 = p_{3,\min} + x\frac{L_2-L_1}{L_3-L_2}
\end{align}
\end{subequations}
with $p_{2,\max}=(c_1L_3-c_2)/(L_3-L_2)$ and $p_{3,\min}=(c_2-c_1L_2)/(L_3-L_2)$.  If we define $f(L_i)$ to be the contribution of a single dipole moment to the extinction per unit volume
\begin{align}
f(L_i) = \Im\left( \frac{1}{L_i - \xi} \right) = \frac{\xi_i}{(L_i-\xi_r)^2 + \xi_i^2}
\end{align} 
then the total extinction from the three eigenvalues is
\begin{align}
\frac{\sigma_{\textrm{ext}}}{V} = &xf(L_1) + \left( p_{2,\max}-x\frac{\Delta_{31}}{\Delta_{32}} \right) f(L_2) \nonumber \\
&+ \left( p_{3,\min}+x\frac{\Delta_{21}}{\Delta_{32}} \right) f(L_3)
\end{align}
where $\Delta_{ij} = L_i-L_j$.  The derivative of the extinction with respect to $x$,
\begin{align}
\frac{\partial}{\partial x} \left( \frac{\sigma_{\textrm{ext}}}{V} \right) = f(L_1) + \frac{ \Delta_{21}f(L_3) - \Delta_{31}f(L_2) }{\Delta_{32}}
\end{align}
is a \emph{constant} with respect to $x$, meaning the optimal $x$  is obtained on the boundary: either $x=0$ ($p_1=0$), or $x=x_{max}$, in which case $p_2=0$.  Depending on the sign of $\Delta_{21}f(L_3)-\Delta_{31}f(L_2)$, either $L_1$ or $L_2$ could be optimal; mixing dipole moments among both cannot be. The math works out identically in the case that both $L_2$ and $L_3 \geq 1/3$, choosing $L_3$ as the independent variable.

Therefore, among any three distinct eigenvalues, the dipole moments should be distributed such that one eigenvalue has zero dipole moment.  For sets of any number of distinct eigenvalues, this procedure can be repeated until there are only two remaining eigenvalues.  The optimal distribution must have only one or two distinct eigenvalues.

We turn to find the optimal $L_1$, $L_2$, $p_1$, and $p_2$.  First we prove an extra condition that will be useful.  If $\xi_r \leq 1/3$, the optimal distribution \emph{cannot} have any $L_i<\xi_r$.  This is because we can trivially improve the figure of merit while satisfying the sum rules.  Say $L_1<\xi_r$ and $L_2>1/3$.  Then we can infinitesimally increase $L_1$ and decrease $L_2$, such that the sum rule is satisfied, and \emph{both} moves increase the figure of merit (note that $L_2$ can always decrease, as some dipole moment must be distributed to the right of, and not equal to, $1/3$.  If there were no $L_i$ greater than $1/3$, there could be no $L_i$ less than $1/3$, either).  Similarly, if $\xi_r \geq 1/3$, then the optimal distribution cannot have any $L_i>\xi_r$.  So whatever the optimal distribution, the position of the eigenvalues relative to $\xi_r$ must all be the same ($L_i-\xi_r \geq 0$ if $\xi_r<1/3$ and $L_i-\xi_r \leq 0$ if $\xi_r>1/3$).  In the case of $\xi_r=1/3$ this means that only the single $L_1=1/3$ is optimal.

Rather than write down the optimization problem as a constrained optimization in four variables, we solve the unconstrained problem over just $L_1$ and $L_2$:
\begin{align}
\max_{L_1,L_2} f(L_1,L_2) = \frac{p_1(L_1,L_2) \xi_i}{(L_1-\xi_r)^2 + \xi_i^2} + \frac{p_2(L_1,L_2) \xi_i}{(L_2-\xi_r)^2 + \xi_i^2} 
\end{align}
where
\begin{subequations}
\begin{align}
p_1(L_1,L_2) = \frac{3L_2 - 1}{L_2-L_1} \\
p_2(L_1,L_2) = \frac{1-3L_1}{L_2-L_1}
\end{align}
\end{subequations}
and $L_1 \in [0,1/3]$, $L_2 \in [1/3,1]$.  The optimal distribution either has $\partial f /\partial L_1 = 0$ and $\partial f / \partial L_2 = 0$, or lies on the boundary.  The derivatives are:
\begin{subequations}
\begin{align}
\frac{\partial f}{\partial L_1} &= \frac{p_1 \xi_i}{x_2-x_1} \left[ \frac{-2 x_1 (x_2-x_1)}{\left( x_1^2 + \xi_i^2 \right)^2} + \frac{1}{ x_1^2 + \xi_i^2 } - \frac{1}{x_2^2 + \xi_i^2} \right] \\
\frac{\partial f}{\partial L_2} &= \frac{p_2 \xi_i}{x_2-x_1} \left[ \frac{-2 x_2 (x_2-x_1) }{\left( x_2^2 + \xi_i^2 \right)^2} + \frac{1}{ x_1^2 + \xi_i^2 } - \frac{1}{x_2^2 + \xi_i^2} \right]
\end{align}
\end{subequations}
where $x_i=L_i-\xi_r$.  It will also be useful to have the second derivatives $\partial^2 f / \partial x_i^2$ (the mixed derivative won't be necessary), which we write here:
\begin{subequations}
\begin{align}
\frac{\partial^2 f}{\partial x_1^2} = \frac{2\xi_i \left[ x_1^2 (x_1+3x_2) - (3x_1+x_2)\xi_i^2 \right]\left(-1+3x_2+3\xi_r\right)}{\left( x_1^2 + \xi_i^2 \right)^3 \left(x_2^2 + \xi_i^2 \right)} \label{eq:d2fdx1}\\
\frac{\partial^2 f}{\partial x_2^2} = \frac{2\xi_i \left[ x_2^2 (x_2+3x_1) - (3x_2+x_1)\xi_i^2 \right]\left(-1+3x_1+3\xi_r\right)}{\left( x_2^2 + \xi_i^2 \right)^3 \left(x_1^2 + \xi_i^2 \right)} \label{eq:d2fdx2}
\end{align}
\end{subequations}
One possible solution for both first derivatives to equal zero is $x_1=x_2$ (i.e. $L_1=L_2=1/3$); however, that is a boundary value we will show later can be ignored (note that the derivative does not blow up if $x_2=x_1$, if you carefully evaluate the two right-hand-side terms).  This has the additional benefit that $p_1$ and $p_2$ can be safely divided out of the problem.  For the first derivative to equal zero, we find the simple condition
\begin{align}
x_2 = \frac{\xi_i^2-x_1^2}{2x_1}
\label{eq:derivL1Soln}
\end{align}
The second derivative is identical to the first, but with $x_2 \leftrightarrow x_1$, so we also have
\begin{align}
x_1 = \frac{\xi_i^2-x_2^2}{2x_2}
\label{eq:derivL2Soln}
\end{align}
The solutions for both $x_1$ and $x_2$ cannot be simultaneously satisfied unless $x_2=x_1$, which we have explicity disallowed.  Thus the optimal distribution \emph{has at least one eigenvalue on the boundary}.  We note that if one of the $L_i=1/3$, then both of the $L_i=1/3$, in order for $\left\langle L_n \right\rangle = 1/3$, further reducing the space of possible values.  We now treat four separate cases:

\underline{Case 1: $0 < \xi_r \leq 1/3$}  
In this case we know $L_1 \geq \xi_r$, which further disallows the boundary value $L_1=0$.  Thus the only possible solutions are $L_1=L_2=1/3$, or $L_2=1$ and $L_1 \in [\xi_r,1/3)$.  If $L_2=1$, only the single derivative $\partial f / \partial L_1$ must equal zero, which previously yielded Eq.~(\ref{eq:derivL1Soln}).  Solving for $x_1$, and discarding the negative solution ($x_1\geq 0$ since $L_1\geq \xi_r$) finds the optimal value $x_1^*$
\begin{align}
x_1^* = \sqrt{x_2^2 + \xi_i^2} - x_2 = \sqrt{(1-\xi_r)^2 + \xi_i^2} - 1 + \xi_r
\label{eq:optX1Case1}
\end{align}
To fulfill the condition $L_1 \leq 1/3$ ($x_1 \leq 1/3-\xi_r$), $\xi_r$ and $\xi_i$ must satisfy:
\begin{align}
\sqrt{(1-\xi_r)^2 + \xi_i^2} -1 + \xi_r \leq 1/3 - \xi_r
\end{align}
Solving, one finds
\begin{align}
3\xi_i^2 \leq \left( \xi_r - 7/9 \right) \left( \xi_r - 1/3 \right)
\label{eq:Case1Domain}
\end{align}
which is the equation for a hyperbola in the $(\xi_r,\xi_i)$ plane.  Since $\xi_r \leq 1/3$, only half of the hyperbola is relevant.

For $(\xi_r,\xi_i)$ that satisfy Eq.~(\ref{eq:Case1Domain}), there remains the question of whether it is a maximum, and if it is a global maximum.  We can answer both in the affirmative.

In Eq.~(\ref{eq:d2fdx1}) for the second derivative with respect to $x_1$, we can see immediately that the right-most term in the numerator, $-1+3x_2+3\xi_r=-1+3L_2>0$.  Then by inserting Eq.~(\ref{eq:optX1Case1}) into the term in square brackets in Eq.~(\ref{eq:d2fdx1}), it is straightforward to show that $\partial^2 f / \partial x_1^2 < 0$.  Moreover, $x_1^*$ is the only local optimum for all possible values of $x_1$, which means it must also be the global optimum.

Don't we have to also compare to the boundary value $L_1=L_2=1/3$?  Fortunately, no.  In the preceding argument, in which $L_2$ was fixed at $1$, the boundary $L_1=1/3$ was sufficient, as $(1/3,1)$ is actually equivalent to $(1/3,1/3)$ (for $L_1=1/3$, $p_1=3$ and $p_2=0$, regardless of the value of $L_2$).  Thus, $L_1=L_2=1/3$ was implicitly checked already, and was found not be be optimal, in the case where $x_1^*$ was valid (cf. Eq.~(\ref{eq:Case1Domain})).  When there is no interior optimal point, the optimum must occur at $(1/3,1/3)$, as $(0,1)$ cannot be optimal ($L_1^* \geq \xi_r$).

\underline{Case 2: $\xi_r < 0$}  
Much of the apparatus of the previous case applies here as well, including the optimal value of $x_1^*$, Eq.~(\ref{eq:optX1Case1}).  Now, however, there is an additional condition on $\xi_r$ and $\xi_i$ that restricts $L_1^*>0$ ($x_1^*>-\xi_r$):
\begin{align}
\xi_i^2 \geq 3\xi_r \left( \xi_r - 2/3 \right)
\label{eq:Case2Domain}
\end{align}
The same argument as in the preceding section again shows that when $x_1^*$ is valid, now given by Eqs.~(\ref{eq:Case1Domain},\ref{eq:Case2Domain}), it must be the globally optimal point.  When it is not valid, however, we must now decide whether $(1/3,1/3)$ is optimal, or whether $(0,1)$ is optimal.  We can write out
\begin{align}
f&\left(\frac{1}{3},\frac{1}{3}\right) - f(0,1) \nonumber \\
&= \frac{2 \xi_i \left( -1 + 3\xi_i^2 + 8\xi_r - 9\xi_r^2 \right)}{9\left(\xi_i^2+\xi_r^2\right) \left(\xi_i^2 + (1-\xi_r)^2\right) \left(\xi_i^2 + (1/3-\xi_r)^2\right)}\
\label{eq:bndDiffCase2}
\end{align}
In the case where Eq.~(\ref{eq:Case2Domain}) is invalidated, it is easy to insert the opposite condition into Eq.~(\ref{eq:bndDiffCase2}) and verify that the numerator, $2\xi_i(2\xi_r-1) < 0$, meaning that $(0,1)$ are the optimal $(L_1,L_2)$.  Conversely, when Eq.~(\ref{eq:Case1Domain}) is invalidated, one can verify that $f(1/3,1/3)>f(0,1)$, such that $(1/3,1/3)$ is optimal.

The next two cases will be highly symmetric with the previous two, so we will just highlight the differences.

\underline{Case 3: $1/3 < \xi_r \leq 1$}  
The difference between this case and the first one is that we now know $L_2\leq \xi_r$, disallowing the $L_2=1$ boundary from being optimal but allowing $L_1=0$ to potentially be optimal.  For $L_1=0$ ($x_1=-\xi_r$), setting $\partial f/\partial x_2=0$ yields the optimal $x_2$:
\begin{align}
x_2^* = -\sqrt{x_1^2+\xi_i^2} - x_1 = \xi_r - \sqrt{\xi_r^2+\xi_i^2}
\end{align}
where we have taken the negative solution because $x_1$,$x_2 \leq 0$.  The condition for $L_2 \geq 1/3$ limits the possible $(\xi_r,\xi_i)$:
\begin{align}
\xi_i^2 \leq 3\left(\xi_r-1/9\right)\left(\xi_r-1/3\right)
\label{eq:Case3Domain}
\end{align}
which is half of a hyperbola in the $\xi$ plane, for $\xi_r>1/3$.  By exactly the same arguments as in the first case, when Eq.~(\ref{eq:Case3Domain}) is satisfied, $x_2^*$ must be globally optimal.  When it is not, $(1/3,1/3)$ must be the optimal value.

\begin{figure}
\includegraphics[width=\linewidth]{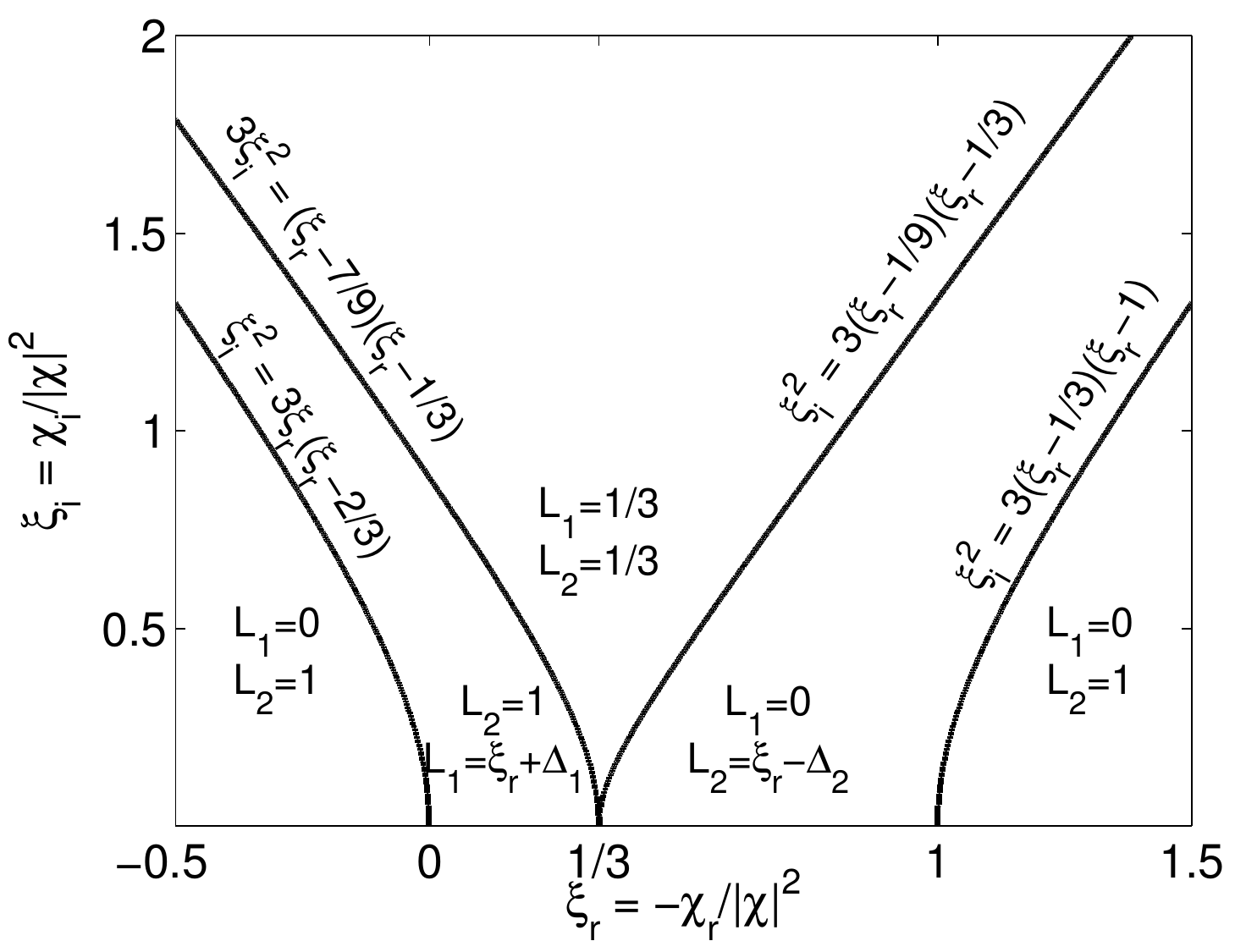}
\caption{\label{figS1}Demarcation of the optimal depolarization factors in the $(\xi_r,\xi_i)$ plane.  The solution presented in the main text is the exact solution for $\xi_i=0$.}
\end{figure}

\begin{figure}
\includegraphics[width=\linewidth]{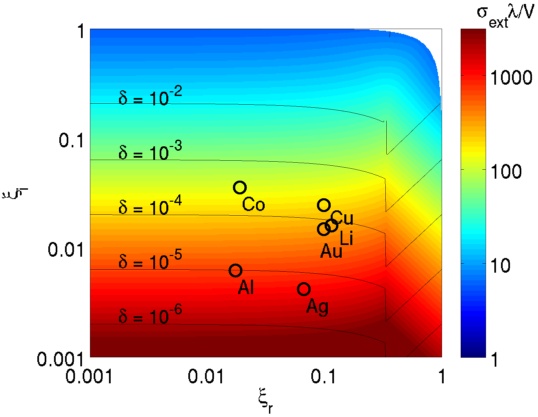}
\caption{\label{figS2}Image: fundamental limits to $\sigma_{\textrm{ext}}\lambda/V$ as a function of $\xi$ in the complex plane.  Contour lines: relative error $\delta$ between  simplified solution, Eq.~(7) in the main text, and the exact solution, Eq.~(\ref{eq:exactSoln}).  For most materials, the error is very small, and the simplified solution is sufficient.}
\end{figure}

\underline{Case 4: $\xi_r > 1$}  
Allowing $\xi_r>1$ introduces the extra potentially optimal boundary $L_2=1$, and it introduces the further condition on $(\xi_r,\xi_i)$ for $x_2^*$ to be valid:
\begin{align}
\xi_i^2 \geq 3\left( \xi_r - 1/3 \right) \left( \xi_r - 1 \right)
\label{eq:Case4Domain}
\end{align}
If this further condition is met, then $x_2^*$ is globally optimal.  Otherwise, we again compare $f(1/3,1/3)$ to $f(0,1)$, through Eq.~(\ref{eq:bndDiffCase2}), when the conditions are not met.  One finds that when Eq.~(\ref{eq:Case4Domain}) is not met, $(0,1)$ is optimal, and when Eq.~(\ref{eq:Case3Domain}) is not met, $(1/3,1/3)$ is optimal.

We can collate the results of the four cases into a single tedious but exact analytical representation for the optimal depolarization factors, $(L_1,L_2)_{\text{opt}}$, for any possible material:
\begin{widetext}
\begin{align}
(L_1,L_2)_{\text{opt}} = 
\begin{cases}
(0,1) & \xi_r<0 \text{; } \xi_i^2 < 3\xi_r\left(\xi_r-\frac{2}{3}\right) \text{ or,}\\
 & \xi_r > 1 \text{; } \xi_i^2 < 3\left(\xi_r-\frac{1}{9}\right) \left(\xi_r-\frac{1}{3}\right) \\
(\xi_r+\Delta_1,1) & \xi_r < 0 \text{; } 3\xi_r\left(\xi_r-\frac{2}{3}\right) \leq \xi_i^2 \leq \frac{1}{3} \left(\xi_r-\frac{7}{9}\right) \left(\xi_r-\frac{1}{3}\right) \text{ or,} \\
 & 0 \leq \xi_r \leq \frac{1}{3} \text{; } \xi_i^2 \leq \frac{1}{3} \left(\xi_r-\frac{7}{9}\right) \left(\xi_r-\frac{1}{3}\right)\\
(0,\xi_r-\Delta_2) & \frac{1}{3} < \xi_r \leq 1 \text{; } \xi_i^2 \leq 3 \left(\xi_r-\frac{1}{9}\right) \left(\xi_r-\frac{1}{3}\right) \text{ or,} \\
 & \xi_r > 1 \text{; } 3\left(\xi_r-\frac{1}{3}\right) \left(\xi_r-1\right) \leq \xi_i^2 \leq 3 \left(\xi_r-\frac{1}{9}\right) \left(\xi_r-\frac{1}{3}\right) \\
\left(\frac{1}{3},\frac{1}{3}\right) & \xi_r \leq \frac{1}{3} \text{; } \xi_i^2 > \frac{1}{3} \left(\xi_r-\frac{7}{9}\right) \left(\xi_r-\frac{1}{3}\right) \text{ or,} \\
 & \xi_r > \frac{1}{3} \text{; } \xi_i^2 > 3 \left(\xi_r-\frac{1}{9}\right)\left(\xi_r-\frac{1}{3}\right)
\end{cases}
\label{eq:exactSoln}
\end{align}
\end{widetext}
where 
\begin{subequations}
\label{eq:Delta12}
\begin{align}
\Delta_1 &= \left( 1 - \xi_r \right) \left( \sqrt{1+\frac{\xi_i^2}{(1-\xi_r)^2}} - 1 \right) \\
\Delta_2 &= \xi_r \left( \sqrt{1+\frac{\xi_i^2}{\xi_r^2}} - 1 \right)
\end{align}
\end{subequations}
The optimal depolarization factors are mapped out in Fig.~\ref{figS1}, which carefully delineates each of the regions.

$\Delta_1 = \Delta_2 = 0$ in the case where $\xi_i=0$, yielding the $(\xi_r,1)$ or $(0,\xi_r)$ optimal values that are used in the main text.  We can see why those values, although not exactly correct when $\xi_i>0$, are very accurate in most cases.  Because the optimal eigenvalues must lie on the boundary, one of the optimal values will usually be exactly correct (e.g. $L_2$ in the case $0\leq \xi_r \leq 1/3$), with the other value off only by a small factor given by Eq.~(\ref{eq:Delta12}).  Fig.~\ref{figS2} shows the very small error between Eq.~(7) and the exact bounds for most materials.

It is instructive to write out the depolarization factors of the optimal spheroids mentioned in the main text.  The optimal depolarization factors of an infinitely flat disk, a sphere, and an infinitely long cylinder (equivalent to an infinitely long ``needle'' spheroid), written in our notation with degeneracies included, are:
\begin{align}
(L_1,L_2)_{\text{spheroid}} = 
\begin{cases}
(0,1) & \text{infinite oblate disk} \\
(0,\frac{1}{2}) & \text{infinite circular cylinder} \\
\left(\frac{1}{3},\frac{1}{3}\right) & \text{sphere}
\end{cases}
\end{align}
One can see from Fig.~\ref{figS1} that the space of theoretically possible materials is largely covered by either the solution $(0,1)$ or the solution $(1/3,1/3)$, for which either flat disks or spheres reach the optimum.  There is an additional curve along which the optimal solution is $(0,1/2)$, for which the cylinder is optimal.  Thus we see why spheroids perform so well.  One should note, as seen in Fig.~\ref{figS2}, that for most materials $0<\xi_r<1/3$ and $\xi_i<0.1$, for which spheroids do not reach the upper bound and some improvement beyond spheroids is possible.

\section{Short proofs of the sum rules}
We provide a unified framework to derive both sum rules, utilizing work by Ouyang and Isaacson \cite{Ouyang1989} and Fuchs \cite{Fuchs1976}, while also introducing the concept of ``resolution of the identity,'' which is well-known in quantum mechanics but appears to not have been recognized explicitly in previous work on integral equation formulations of the scattering problem.  The derivation here of the second sum rule avoids the continuous-to-discrete-to-continuous transformations of \cite{Fuchs1976}, potentially simplifying the proof.  

Following \cite{Ouyang1989} we define the linear operator 
\begin{align*}
B[\sigma](x) &= \int_S \frac{1}{|x-x'|}\sigma(x')d^2 x'
\end{align*}
which is positive definite and symmetric, such that it has a unique Cholesky decomposition
\begin{align}
B = U^T U
\end{align}
where $U$ is upper triangular.  So-called ``Plemelj symmetrization'' yields
\begin{align}
K^{T} B = B K
\end{align}
showing that $BK$ is symmetric.  We define another symmetric operator
\begin{align*}
D &= (U^{-1})^T B K U^{-1} \\
  &= U K U^{-1}
\end{align*}
which has the same eigenvalue spectrum as $K$.  Defining $s_n$ to be the eigenvectors of $D$ ($Ds_n = \lambda_n s_n$), there is a simple relationship between $s_n$ and $\sigma_n$:
\begin{align}
\sigma_n = U^{-1} s_n
\end{align}
$D$ is complete, leading to the resolution of the identity we need:
\begin{align}
\delta(x-x') &= \sum_n s_n(x') s_n(x) \nonumber \\
			 &= \sum_n \sigma_n(x') U^{T} U \sigma_n(x) \nonumber \\
			 &= \sum_n \sigma_n(x') B \sigma_n(x) \nonumber \\
			 &= \sum_n \tau_n(x') \sigma_n(x)
\end{align}
This identity yields the two sum rules.  For the first, we simply sum over the dipole strengths of every mode
\begin{align}
\sum_n p_{n,\alpha\beta} &= \sum_n \frac{ \langle \tau_n, \hat{\mathbf{n}} \cdot \hat{\mathbf{\alpha}} \rangle \langle \sigma_n, \mathbf{r} \cdot \hat{\mathbf{\beta}} \rangle }{V} \nonumber \\
&= \frac{1}{V} \int_S \int_S n_{\alpha}(x) \sum_n \tau_n(x) \sigma_n(x') x'_{\beta} d^2xd^2x' \nonumber \\
&= \frac{1}{V} \int_S n_{\alpha} x_{\beta} d^2 x \nonumber \\
&= \delta_{\alpha\beta}\frac{1}{V} \int_V \nabla \cdot x_{\alpha} d^3 x \nonumber \\
&= \delta_{\alpha\beta}
\end{align}
For the second sum rule, we use the crucial formula
\begin{align}
\sum_n \tau_n(x')\lambda_n\sigma_n(x) = K(x,x')
\end{align}
which can be proven by taking
\begin{align}
\sum_n \tau_n(x') \lambda_n \sigma_n(x) &= \sum_n \tau_n(x') \int_S K(x,y) \sigma_n(y) d^2y \nonumber \\
&= \int_S d^2y \, K(x,y) \sum_n \tau_n(x') \sigma_n(y)  \nonumber \\
&= K(x,x')
\end{align}
where in the final step we have again used the resolution of the identity.
Now we can write
\begin{align}
\sum_{n,\alpha} \lambda_n p_{n,\alpha\alpha} &= \frac{1}{V} \sum_{\alpha}\int_S d^2 x \int_S d^2 x' \, n_{\alpha}(x) x'_{\alpha} \sum_n \tau_n(x) \lambda_n \sigma_n(x') \nonumber \\
&= \frac{1}{V} \int_S d^2 x \int_S d^2 x' \sum_{\alpha} x'_{\alpha} K(x',x) n_{\alpha}(x) \nonumber \\
&= \frac{1}{V} \int_S \int_S \, x \cdot n(x') \left(-n(x) \cdot \nabla_{x} \frac{1}{|x-x'|} \right) d^2 x d^2 x'
\label{eq:sumLambdaP}
\end{align}
The final integral in Eqn.~\ref{eq:sumLambdaP} is worked out in \cite{Fuchs1976} and shown to be equal to $-1/2$, for either a single particle or a collection of particles.  Thus we have
\begin{align}
\sum_{n,\alpha} \lambda_n p_{n,\alpha\alpha} = -\frac{1}{2}
\end{align}
or, in terms of the depolarization factor
\begin{align}
\left\langle L_n \right\rangle = \frac{ \sum_{n,\alpha} L_n p_{n,\alpha\alpha} }{ \sum_{n,\alpha} p_{n,\alpha\alpha} } = \frac{1}{3}
\end{align}
